\documentclass[global,twocolumn,referee]{svjour}
\usepackage{latexsym}
\usepackage{graphics}
\usepackage{graphicx}
\usepackage{cite}
\usepackage{color}
\usepackage{bm}
\usepackage{float}
\usepackage{amssymb}
\usepackage{amsmath}
\journalname{my journal}
\begin{document}
\title{A tunable Doppler-free dichroic lock for laser frequency stabilization}
\subtitle{}
\author{Vivek Singh \and V. B. Tiwari \and S. R. Mishra and H. S. Rawat
}                     
%
%
\institute{Laser Physics Applications Section, Raja Ramanna Center for Advanced Technology, Indore-452013, India.}
\date{Received: date / Revised version: date}
%
\maketitle
\begin{abstract}
We propose and demonstrate a laser frequency stabilization scheme which generates a dispersion-like tunable Doppler-free dichroic lock (TDFDL) signal. This signal offers a wide tuning range for lock point (i.e. zero-crossing) without compromising on the slope of the locking signal. The method involves measurement of magnetically induced dichroism in an atomic vapour for a weak probe laser beam in presence of a counter propagating strong  pump laser beam. A simple model is presented to explain the basic principles of this method to generate the TDFDL signal. The spectral shift in the locking signal is achieved by tuning the frequency of the pump beam. The TDFDL signal is shown to be useful for locking the frequency of a cooling laser used for magneto-optcal trap (MOT) for $^{87}Rb$ atoms.
\end{abstract}

\section{Introduction}
\label{intro}
Laser frequency stabilization is an essential requirement for various applications including atom cooling\cite{chu}, high resolution spectroscopy\cite{CE}, precision measurements\cite{JE} etc. A setup for laser cooling and trapping of atoms requires several lasers which are actively frequency stabilized and locked at few line-widths detuned from the peak of atomic absorption. In the laser cooling experiments, the active frequency stabilization is achieved by generating a reference signal which is based on absorption profile of atom around the resonance. The reference locking signal can be either Doppler-broadened with wide tuning range or Doppler-free with comparatively much steeper slope but with limited tuning range. The most commonly used technique for frequency locking involves locking to the side of the Doppler free peak generated from the saturated absorption spectroscopy (SAS)\cite{Demt, Mac}. The locking of the laser frequency to the peak of an atomic reference signal such as in frequency modulated spectroscopy (FMS)\cite{Mud} provides high signal to noise ratio (SNR) but needs dithering of laser frequency and a phase sensitive detection system. A suitable dispersion-like reference signal for peak locking can be generated using polarizing spectroscopy (PS)\cite{Wie, Pear, Tiwari, VBT, Kale} without frequency modulation and phase sensitive detection.  Along with these techniques, there are also methods for laser frequency stabilization which employ PS with external magnetic field\cite{Krzemien, Sho}. However, the techniques involving polarization spectroscopy are sensitive to the surrounding stray magnetic field\cite{Gawlik}. An alternative technique which can generate narrow dispersion like reference signal exploits circular dichroism of an atomic vapour in the presence of a magnetic field. In this  technique, the difference of Zeeman shifted saturated absorption signals generate the Doppler free dichroic lock (DFDL)signal\cite{Shim, Wasik, Harris, Dian, Hongli, Miyabe, Noh, Pahwa, Marin, Petelski}. This technique requires only low magnetic field for its operation and is less sensitive to the surrounding stray magnetic field. However, the tuning range of this technique is limited to few natural linewidths of the transition. The tuning range can be increased by electronically adding an offset voltage to the signal or by optically changing the quarter wave plate axis relative to the polarisation beam splitter axis\cite{Wasik}. The locking range can also be extended by either using Doppler-broadened version of this technique known as dichroic atomic vapor laser lock (DAVLL)\cite{Corwin, Alfred, Imanishi, Beverini} signal or by locking at positive as well as negative slopes of DFDL signal \cite{Wasik}. Another method involves increasing the magnetic field \cite{VB} to extend the locking range of the DFDL signal. This frequency tuning is often achieved by compromising the slope of the signal which is a crucial parameter for frequency locking. A recently reported study on DFDL spectroscopy of D-2 line of $^{87}Rb$ includes coherence for calculation of line-shape of real atomic system \cite{hrNoh}. \\
Here, we present a technique which generates a narrow tunable DFDL (TDFDL) signal without compromising with the slope of the signal. The frequency tuning of this dispersion like TDFDL signal is achieved by varying the pump laser frequency using AOMs. We have studied the dependence of this frequency locking signal's amplitude and the slope on various experimental parameters such as the pump beam power and the magnetic field. The locking performance of an extended cavity diode laser (ECDL) system is investigated by measuring change in the number of cold atoms in a magneto-optical trap (MOT). A simple theoretical model to explain the generation of TDFDL signal is also presented.
\section{Theoretical analysis}
We adopt a simple two-level system involving hyperfine energy levels to explain the generation of TDFDL signal. A linearly polarized weak probe beam is overlapped by a counter-propagating strong pump beam in an atomic vapor cell placed in a weak magnetic field. The propagation of the probe beam is along the axis of magnetic field. The linearly polarized light can be resolved into two counter-rotating circular components $\sigma_{\pm}$ using quarter wave plate and polarizing beam splitter. In the absence of magnetic field (B=0), different $m_{F'}$ states are degenerate and hence, $\sigma_{+}$ and $\sigma_{-}$ transitions overlap. When a small magnetic field (few Gauss) is applied, the degeneracy is lifted and medium becomes dichroic due to the displacement of $\sigma_{+}$ and $\sigma_{-}$ transitions. As a result, the pair of spectrally shifted saturated absorption spectroscopy (SAS) signals on subtraction generate the Doppler free dispersion like signal\cite{VB}. For atoms moving with velocity $\vec{v}$, the Doppler shifted laser frequency in the frame of moving atom becomes $\omega^{'}=\omega_{0}- \vec{k}.\vec{v}$. When the laser beam passes through a Rb atomic sample with a Maxwell-Boltzmann velocity distribution, only those atoms which fall within the homogeneous linewidth $\Gamma$ around the center frequency $\omega_{0}$ of Rb atom at rest can significantly contribute to the absorption.\\
\begin{figure*}
\centering
\includegraphics[width=10.0 cm]{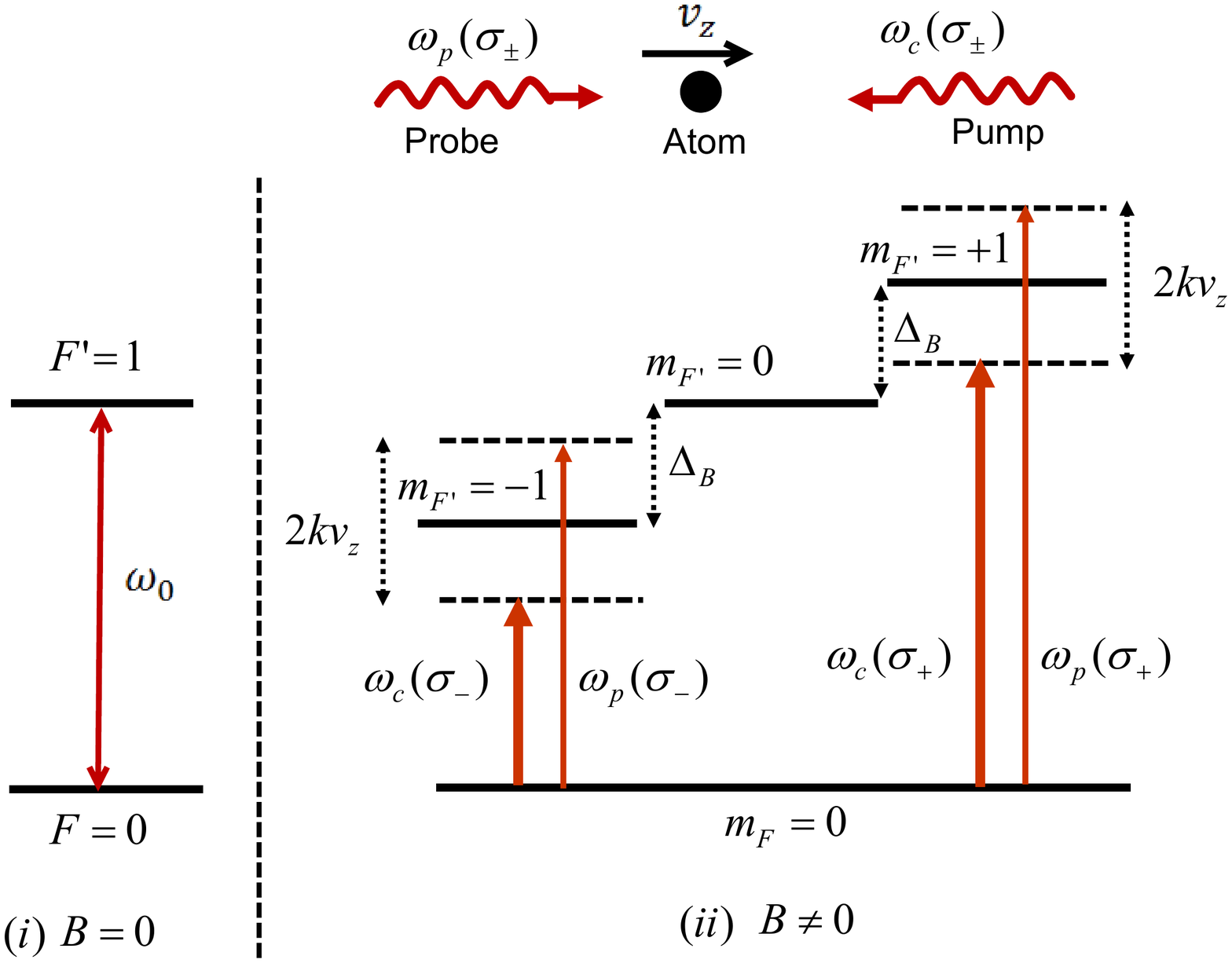}
\vspace*{0.1cm}       
\caption{Schematic diagram showing resonant interaction of $\sigma_{+}$ and $\sigma_{-}$ polarized pump and the probe beams when a weak magnetic field is applied to an atom having lower level $F=0$ and upper level $F'=1$ is in motion.}
\label{fig:1}       
\end{figure*}
If the pump beam having frequency ($\omega_{c}$) and  the probe beam having frequency ($\omega_{p}$) are in counter propagating directions, the atoms having velocity components of $v_{z}$ along direction of the probe beam see different values of frequency detunings with the pump and the probe beams in the presence of magnetic field. The $\sigma_{\pm}$ polarized pump and the probe beams can interact resonantly with this group of atoms if,
\begin{eqnarray}
\omega_{p}(\sigma_{\pm})-kv_{z}=\omega_{0}\pm\Delta_{B}
\end{eqnarray}     
\begin{eqnarray}
\omega_{c}(\sigma_{\pm})+kv_{z}=\omega_{0}\pm\Delta_{B}
\end{eqnarray}     
Using equation (1) and (2), the relation between the probe beam detuning  $\Delta_{p}^\pm =\omega_{p}(\sigma_{\pm})-\omega_{0}$  and the pump beam detuning $\Delta_{c}^\pm =\omega_{c}(\sigma_{\pm})-\omega_{0}$ for resonant interaction is given by
\begin{eqnarray}
 \Delta_{c}^\pm + \Delta_{p}^\pm= \pm 2\Delta_{B}
\end{eqnarray} 
Again using equation (1) and (2), the difference between the pump and the probe beam detunings is given by,
\begin{eqnarray}
\Delta_{cp} = \Delta_{c}^\pm-\Delta_{p}^\pm = -2kv_{z}
\end{eqnarray} 
Using equation (3) and (4), resonance condition for $\sigma_{\pm}$ probe beam in presence of counter propagating pump beam and applied magnetic field can be written as
 \begin{eqnarray}
\Delta_{p}^\pm = -\dfrac{\Delta_{cp}}{2}\pm \Delta_{B}
 \end{eqnarray}
Therefore, the probe absorption signal will depend upon the difference between the pump and the probe beams detunings and the applied magnetic field.\\
In the case of simple model, $F=0 \rightarrow F'=1$, the absorption cross-section for $\sigma_{\pm}$ probe beam for an atom with velocity component $v_{z}$ is given as\cite{Demt}
\begin{eqnarray}
\rho_{\pm}(\Delta_{p}^\pm, v_{z})=\dfrac{\rho_{0}(\Gamma/2)^{2}}{(\Delta_{p}^\pm \mp \Delta_{B}-kv_{z})^{2}+(\Gamma/2)^{2}}
\end{eqnarray}
where $\rho_{0} $ is maximum absorption cross section at $\Delta_{p}^\pm = \pm\Delta_{B}+kv_{z}$, $\Delta_{B} =\mu_{B}Bg_{F'}m_{F'}/\hbar$ (for $F=0 \rightarrow F'=1$ case) is Zeeman splitting of the levels in magnetic field, $k$ is the wave vector of laser light.\\
Due to pump laser beam ($\omega_{c}$), population density in ground state ($F=0$ level) decreases within the velocity interval $dv_{z}=\frac{\Gamma}{k}$, while population density in magnetically shifted upper level ($F'=\pm 1$ level) increases. Therefore, population in the lower level after pump has caused excitation from lower level is given by 
\begin{multline}
\ \Delta N(\Delta_{c}^\pm ,v_{z})= N^{0}(v_{z})[1- \\ 
-g_{1}\dfrac{(\Gamma/2)^{2}}{(\Delta_{c}^+ -\Delta_{B}+kv_{z})^{2}+(\Gamma/2)^{2}} \\ 
-g_{2}\dfrac{(\Gamma/2)^{2}}{(\Delta_{c}^- +\Delta_{B}+kv_{z})^{2}+(\Gamma/2)^{2}}]
\
\end{multline}
where $\Gamma$ is a natural line width for $F=0 \rightarrow F^{'}=1 $ transition, $g_{1}$ and $g_{2}$ are coefficients representing the depth of dips burnt in the distribution by the pump beam and depends on the pump beam intensity used. For our calculations, we have assumed $g=g_{1}=g_{2}$. The population in the ground state with velocity range $v_{z}$ and $v_{z}+dv_{z}$ from where probe makes absorption transition is given as
\begin{eqnarray}
 N^{0}(v_{z})dv_{z}= C exp{[-(v_{z}/v_{p})^2]}dv_{z}
\end{eqnarray}
where $v_{p}$ is most probable velocity and $C$ is a constant depending on number of atoms in gas, density etc..\\
The absorption coefficient for a weak probe laser beam in presence of the pump beam is given by
\begin{eqnarray}
\alpha_{\pm}(\Delta_{p}^\pm, \Delta_{c}^\pm)=\int_{-\infty}^{\infty}\rho_{\pm}(\Delta_{p}^\pm, v_{z})\Delta N(\Delta_{c}^\pm,v_{z})dv_{z}
\end{eqnarray}
The integration in (9) can be performed over velocity and assuming the Doppler width, $\Delta_{D}>>\Gamma$ (where $\Delta_{D}=kv_{p}$) to obtain $\alpha_{\pm}$ values.\\
For low optical density of the sample, the TDFDL signal (represented by $DS$) can be calculated to be
\begin{eqnarray}
DS(\Delta_{p}^\pm, \Delta_{c}^\pm)=\alpha_{+}(\Delta_{p}^+, \Delta_{c}^+)-\alpha_{-}(\Delta_{p}^-, \Delta_{c}^-)
\end{eqnarray}
Using equation (7), (8) and (9), the TDFDL signal is
\begin{multline}
\ DS(\Delta_{p}^\pm, \Delta_{c}^\pm)=\\
C[(D_{+}-D_{-})-g(D_{+}L_{+}-D_{-}L_{-})]
\
\end{multline}
where $D_{\pm}=\dfrac{1}{\sqrt{\pi}}\exp{[-((\Delta_{p}^\pm \mp\Delta_{B})/\Delta_{D})^2]}$\\
$L_{\pm}= \dfrac{\Gamma^{2}}{(\Delta_{p}^\pm +\Delta_{c}^\pm \mp 2\Delta_{B})^{2}+(2\Gamma/2)^{2}}$\\
$L_{\pm}$ can be written in terms of $\Delta_{cp}$ using equation (4),
$L_{\pm}= \dfrac{(\Gamma/2)^{2}}{(\Delta_{p}^\pm +\dfrac{\Delta_{cp}}{2}\mp \Delta_{B})^{2}+(\Gamma/2)^{2}}$\\

The signal consists of two parts: the first part represents a Doppler broadened dichroic signal. The second part appears due to presence of pump beam and it is represented by two Doppler-free absorption coefficients of $\sigma_{+}$ and $\sigma_{-}$ components.\\
\begin{figure*}
\centering
\includegraphics[width=8.0 cm]{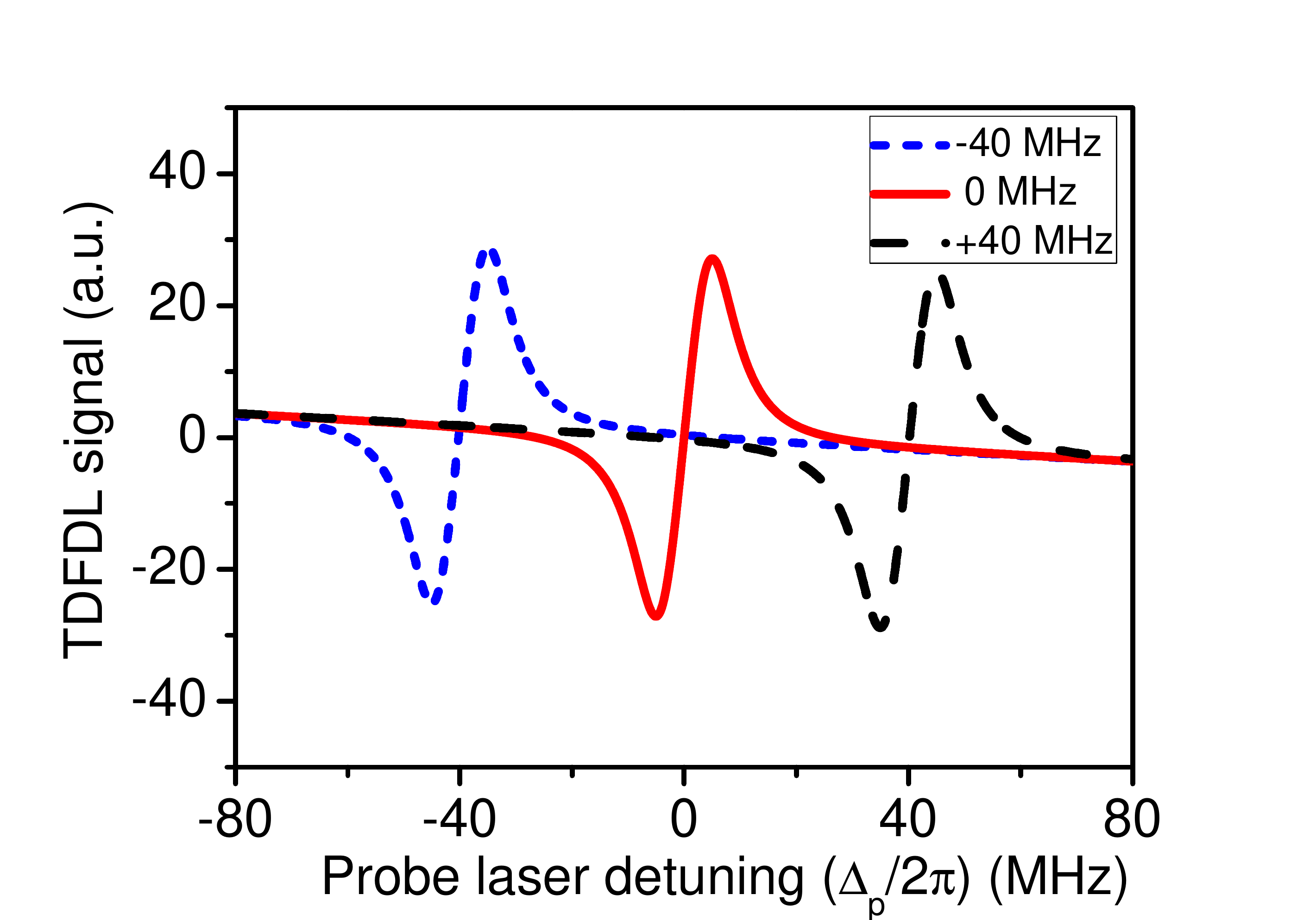}
\vspace*{0.1cm}       
\caption{Calculated TDFDL signals $DS(\Delta_{p}^\pm, \Delta_{c}^\pm)$ (using equation (11)) for the parameter $\Delta_{D}=50\Gamma$, $C=1$, $g=0.04$, $m_{F}=0$, $g_{F'}=2/3$, $m_{F'}=1$, $\Gamma/2\pi= 12$ MHz, $B=4 G$ and $\Delta_{cp}/2\pi$ = +80 MHz, 0 MHz and -80 MHz corresponds to short dashed line, solid line and dashed line respectively.}
\label{fig:2}       
\end{figure*}
Fig. 2 shows the calculated TDFDL signal for different pump laser detuning. As the pump laser frequency is blue detuned with respect to the probe beam, the zero-crossing in dispersion like TDFDL signal will be shifted to red and vice versa. Therefore, the zero crossing in TDFDL signal can be shifted to any frequency depending upon the pump laser detuning. Further, Fig. 3 (a) and (b) shows the calculated amplitude and the slope of the TDFDL signal (12 MHz red detuned) with the axial magnetic field respectively. We find that amplitude (difference of maximum and minimum values) of TDFDL signal initially increases with the applied magnetic field and gets saturated after certain value of the magnetic field. The amplitude of TDFDL signal reaches maximum value when SAS for $\sigma_{+}$ and $\sigma_{-}$ probe beams is just separated by increasing the magnetic field. Further, the slope of the TDFDL signal increases with magnetic field and reaches maximum value at magnetic field of $\sim$ 4 Gauss as shown in fig. 3(b). Slope starts to decrease with further increase of the magnetic field due to increased separation between $\sigma_{+}$ and $\sigma_{-}$ probe laser beam signals.
\begin{figure}
\centering
\begin{minipage}[b]{.4\textwidth}
\includegraphics[width=1.2\linewidth]{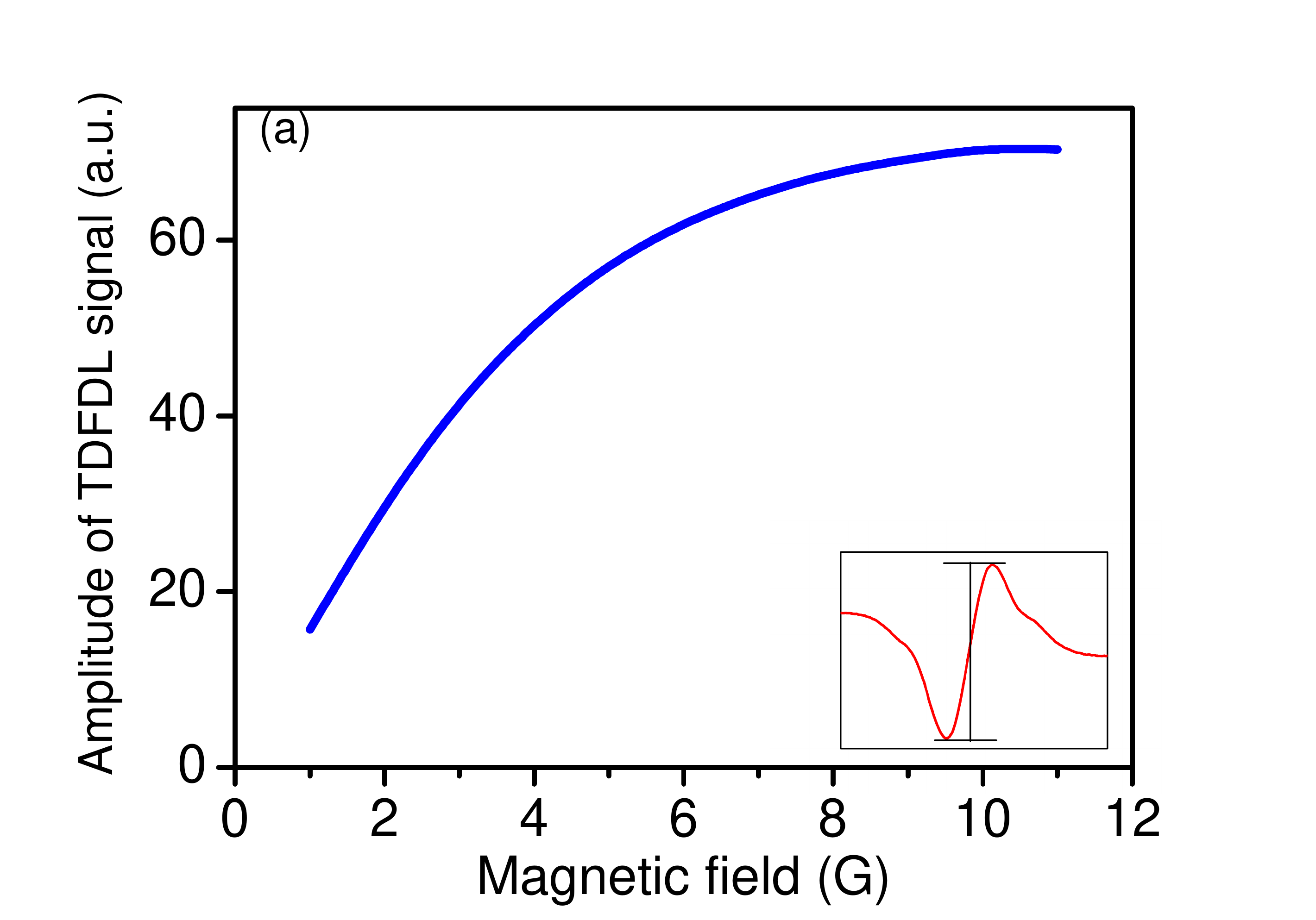}
\end{minipage}\qquad
\begin{minipage}[b]{.4\textwidth}
\includegraphics[width=1.2\linewidth]{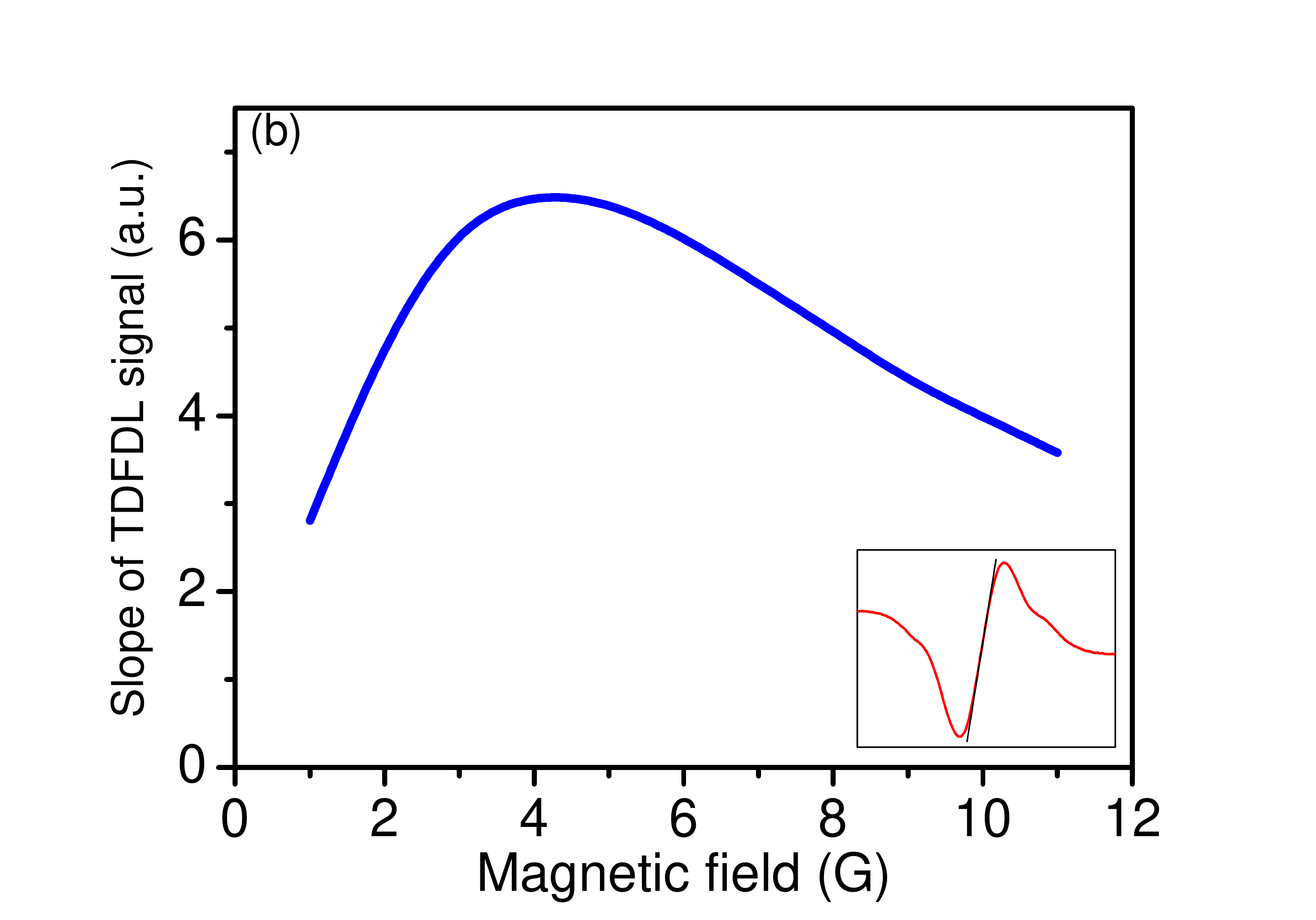}
\end{minipage}
\caption{Calculated dependence of (a) amplitude and (b) Slope of TDFDL signal on magnetic field. $\Delta_{D}=50\Gamma$, $\Gamma/2\pi= 12$ MHz, $C=1$, $g=0.04$, $m_{F}=0$, $g_{F'}=2/3$, $m_{F'}=1$ and $\Delta_{cp}/2\pi= + 24 MHz$.}
\end{figure}
\section{Experimental setup}
The TDFDL experimental setup is shown in Fig. 4. The External cavity diode laser (ECDL) (DL100, Toptica) system operating at 780 nm is used in our experiments. A combination of half wave plate and polarizing beam splitter (PBS) is used to suitably divide laser output beam for its use in TDFDL setup, a reference SAS setup and a magneto-optical trap (MOT) setup. \\
In TDFDL setup, laser light is split in two parts namely the pump laser beam and the probe laser beam using the combination of half wave plate and PBS. The pump laser beam is passed through a setup of two acousto-optical modulators (AOMs), AOM-1 and AOM-2 having frequency range of 60 to 100 MHz. The pump laser beam is initially passed through AOM-1. The first order (+1) diffracted beam of AOM-1 is retro-reflected and passed again through AOM-1 before going to pass through the second AOM i.e. AOM-2 (Fig. 4). In AOM-2, The first order (-1) is made retro-reflected. Thus, we have used  both the AOMs in double pass configuration to avoid deviation of the output beam. The output pump beam is overlapped with the probe laser beam in counter propagating direction inside a 5 cm long Rb vapor cell placed inside the solenoid with crossing of $\sim$ 10 $mrad$. The probe laser beam is monitored using a $\lambda/4$ wave plate and a PBS combination of elements and photo-diodes for detection of $\sigma_{+}$ and $\sigma_{-}$ parts of polarizations (Fig. 4). The difference of two photo-diode signals gives dispersion like signal. The solenoid generates the longitudinal field of $\sim$ 32 G/A. At the Rb vapor cell, the probe and the pump laser beam has a diameter of 3 mm. Neutral density filters are used to vary the pump laser beam power. 

\begin{figure*}
\centering
\includegraphics[width=8.0 cm]{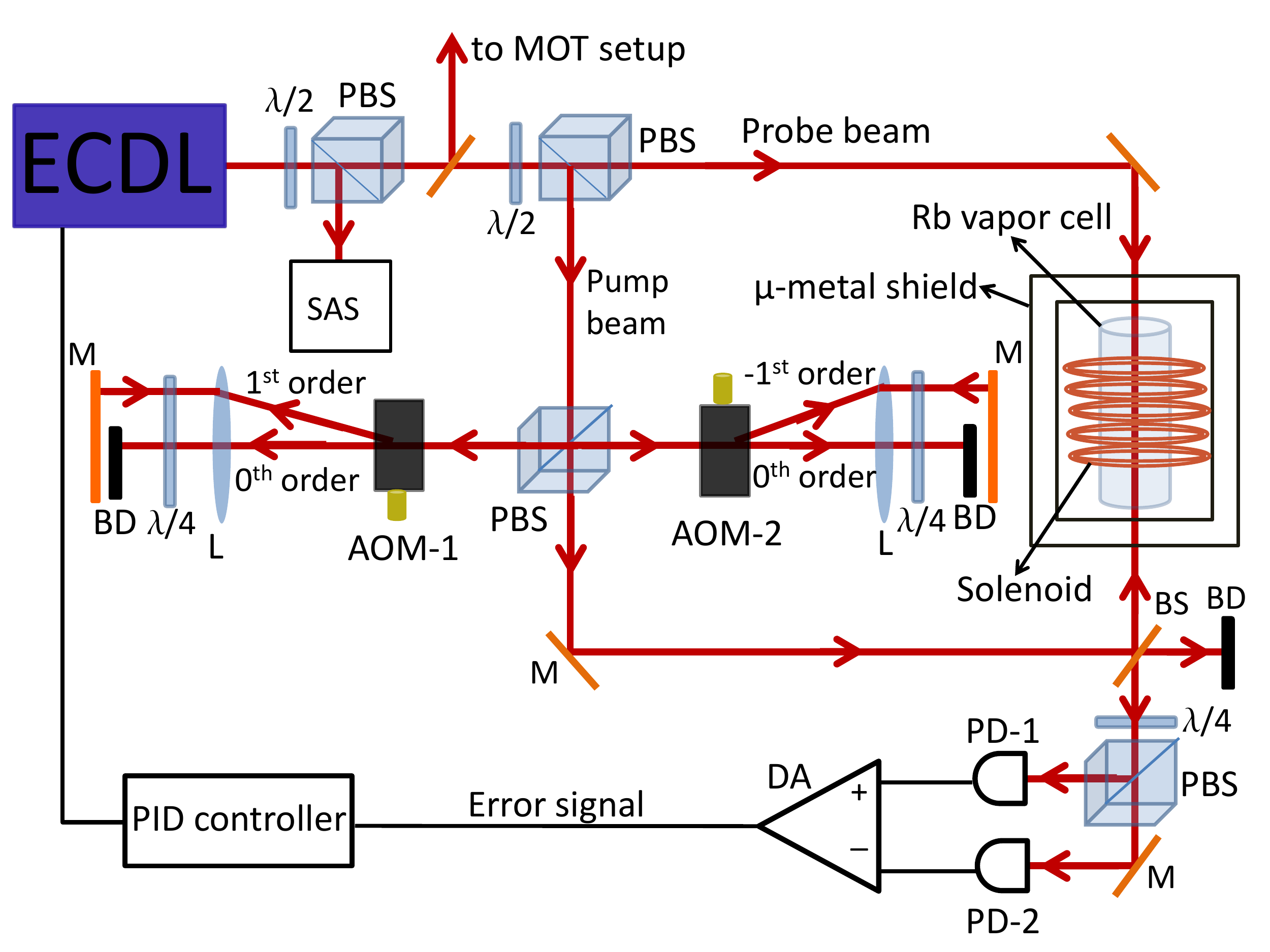}
\vspace*{0.1cm}       
\caption{Schematic diagram of TDFDL spectroscopy setup. ECDL: external cavity diode laser, MOT: magneto-optical trap, PBS: polarizing beam splitter, BD: beam dump, BS: beam splitter, AOM: acousto-optic modulator, DA: differential amplifier, PD: photo-diode, SAS: saturated absorption spectroscopy, M: mirror, L: lens, $\lambda/4$ : quarter wave plate, $\lambda/2$ : half wave plate. Two detectors (PD-1 and PD-2) monitor the absorption of light which has driven $\sigma_{+}$ and $\sigma_{-}$ transitions.}
\label{fig:4}       
\end{figure*}
\section{Results and discussions}
\subsection{Frequency tuning of TDFDL spectra using AOMs}
\label{sec:1} 
The frequency tuning of the TDFDL signal is achieved by varying the frequency of the pump laser beam with the help of two AOMs in double pass configuration as shown in Fig. 4. The pump laser beam after double pass through AOM-1, gets a frequency shift of $2\omega_{1}$, where $\omega_{1}$ is shift in first order diffraction in AOM-1. When this first order diffracted beam passes through AOM-2, the frequency shift is $-2\omega_{2}$, where $\omega_{2}$ is shift in -1st order in AOM-2. Therefore, the frequency of the pump beam is shifted by $\Delta_{AOM}$ ($\Delta_{AOM}=2(\omega_{1}-\omega_{2})=\Delta_{cp}$, relative to the probe beam frequency. Therefore, as we observed, (Fig. 5(a)), there is no spectral shift in the TDFDL signal when AOM-1 and AOM-2 operate at the same frequency. This is because, the pump and the probe laser beam will be interacting with zero velocity group resulting a TDFDL signal as shown in Fig. 5(a) (solid curve). Further, if AOM-1 and AOM-2 operate at frequency 100 MHz and 60 MHz respectively then frequency shift of pump beam ($\Delta_{AOM}$) as compared to the probe beam becomes +80 MHz. Therefore, TDFDL signal will be red shifted (-40 MHz) as shown by short dashed curve in Fig. 5(a). In the same way, TDFDL signal can be shifted to blue detuned (+40 MHz), as shown by dashed curve in Fig. 5(a), if AOM-1 and AOM-2 operate at frequency of 60 MHz and 100 MHz respectively. Thus, observed detuning behaviour of TDFDL signal shown in Fig. 5(a) is in good agreement with our theoretical results shown in Fig. 2.\\ 
We have also observed that the slope and the amplitude of the signal remains nearly unchanged throughout the whole tuning range. Thus, the frequency stability achieved using TDFDL reference signal is expected to be nearly independent of the frequency shifting used. This is qualitatively different from the SAS signal where the slope and the amplitude changes with detuning from the resonance peak. It is shown in Fig. 5b that slope of TDFDL signal is $\sim$ 43 mV/MHz as compared to SAS signal with slope of $\sim$ 6 mV/MHz at -12 MHz detuning. Here, we note that the relative difference in the detuning of the pump and the probe beams can also be introduced using two different lasers as reported in ref.\cite{Kale}. However, in that case the locking performance of one laser may influence the frequency stabilization of the other laser. We note that the tuning of laser frequency using a DFDL setup can also be achieved by changing the frequency of final laser beam using AOMs. But this will be an inefficient method due to loss of power through AOMs used in the final beam. 
\begin{figure}
\centering
\begin{minipage}[b]{.4\textwidth}
 \includegraphics[width=1.2\linewidth]{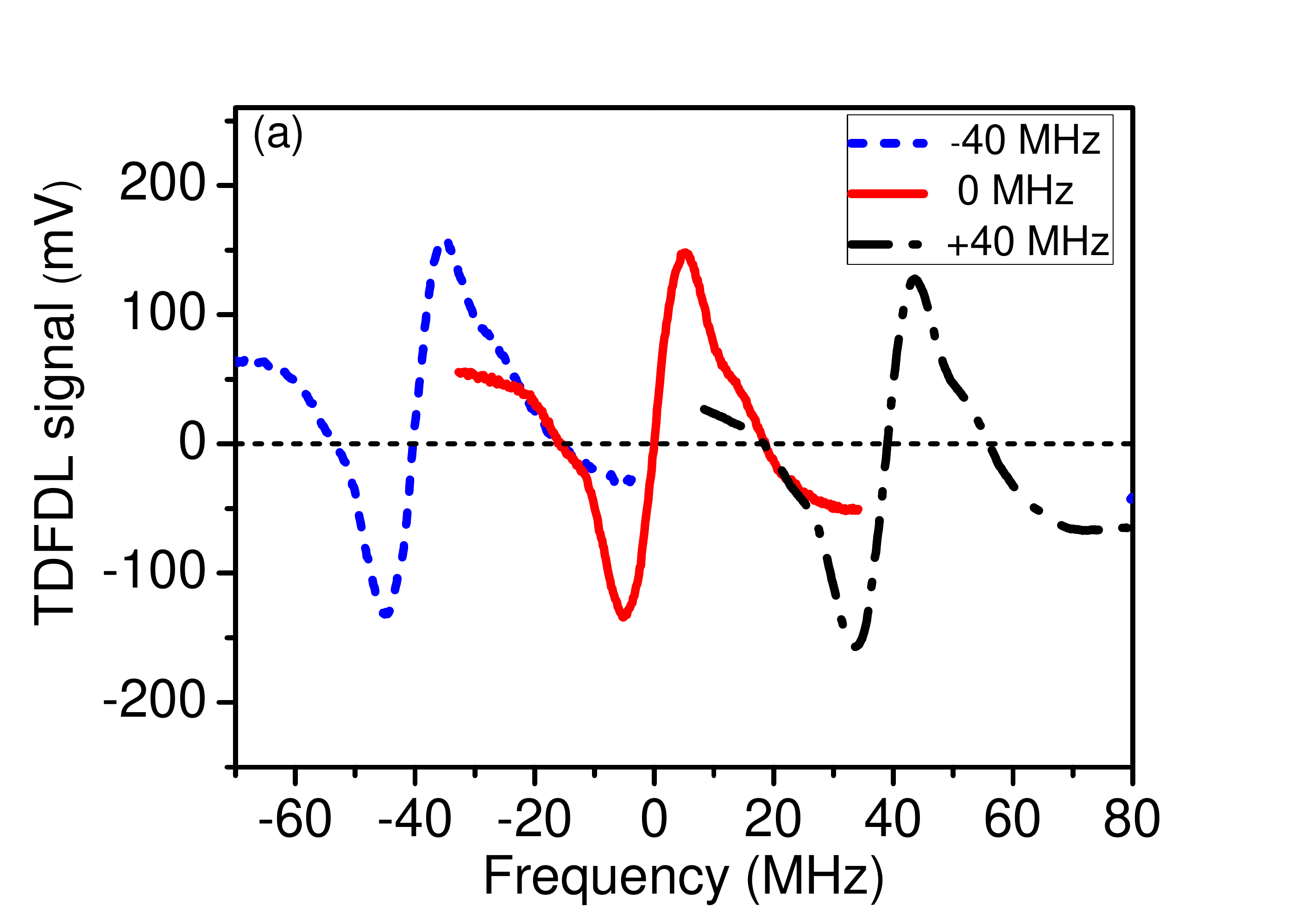}
\end{minipage}\qquad
\begin{minipage}[b]{.4\textwidth}
\includegraphics[width=1.2\linewidth]{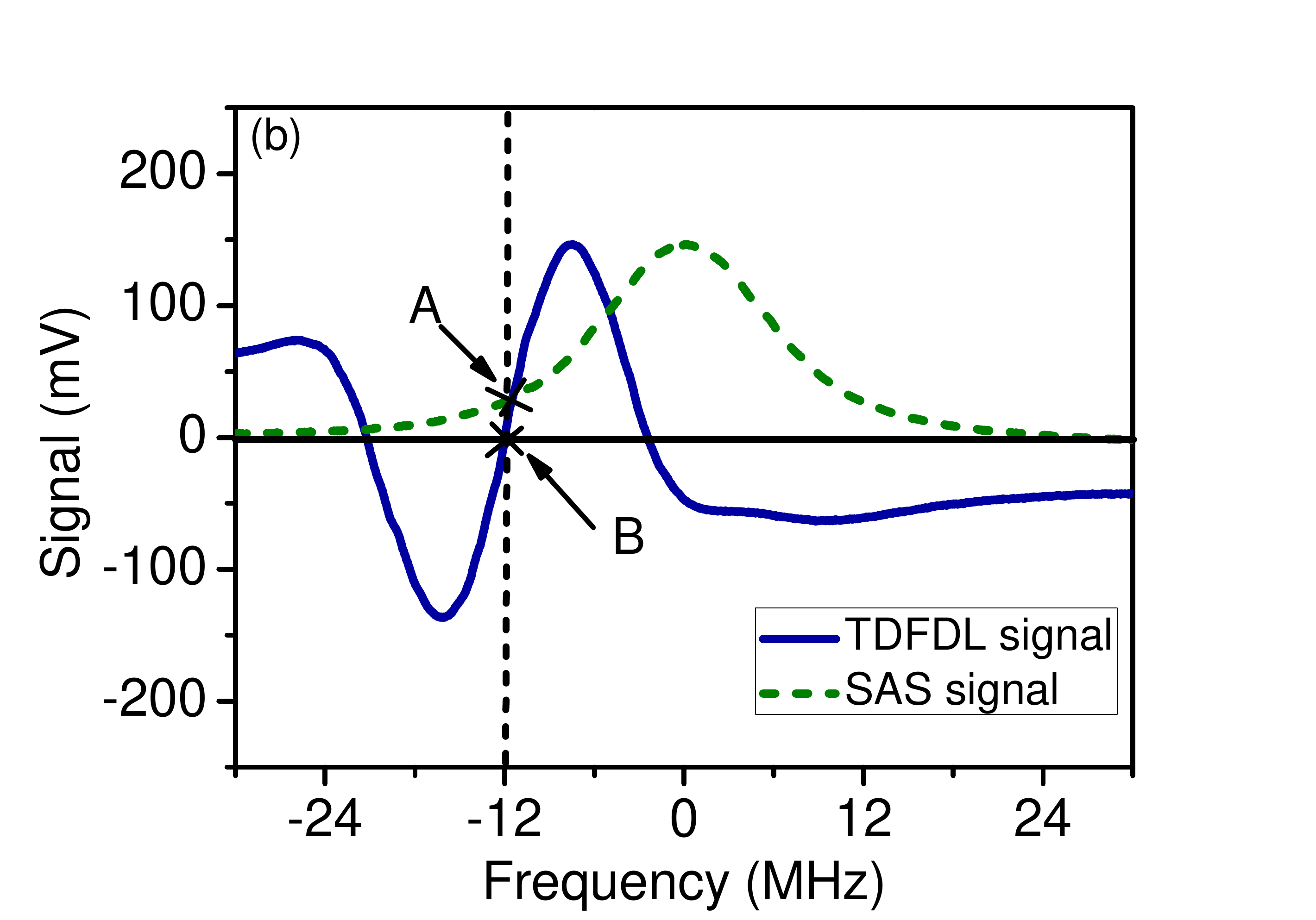}
\end{minipage}
\caption{(a) Observed spectral shift of TDFDL signal for $F=2\rightarrow F'=3$ transition of $^{87}Rb$ atom for $\Delta_{AOM}/2\pi$ = +80 MHz, 0 MHz and -80 MHz (short dashed curve, solid curve and dashed curve respectively). (b) TDFDL signal at -12 MHz detuning with reference to SAS signal. Slope of SAS signal is $\sim$ 6 mV/MHz at point A whereas the slope of TDFDL signal is $\sim$ 43 mV/MHz at point B. pump power = 1.13 mW, Probe power = 140 $\mu$ W, and magnetic field = 9.7 G for the data in figure (b).}
\end{figure}
\subsection{Dependence of TDFDL spectra on magnetic field strength and pump beam power}
\label{sec:2}
This TDFDL signal is utilized to lock the laser frequency at zero-crossing point of the signal. The lock stability of laser is dependent on peak to peak amplitude and slope of the TDFDL signal at the zero crossing position. Therefore, dependence of peak to peak amplitude and slope of the signal on magnetic field strength and the pump beam power is important for the cooling transition  $F=2\rightarrow F'=3$ transition of $^{87}Rb$ atoms.\\ 
\begin{figure}
\centering
\begin{minipage}[b]{.4\textwidth}
 \includegraphics[width=1.2\linewidth]{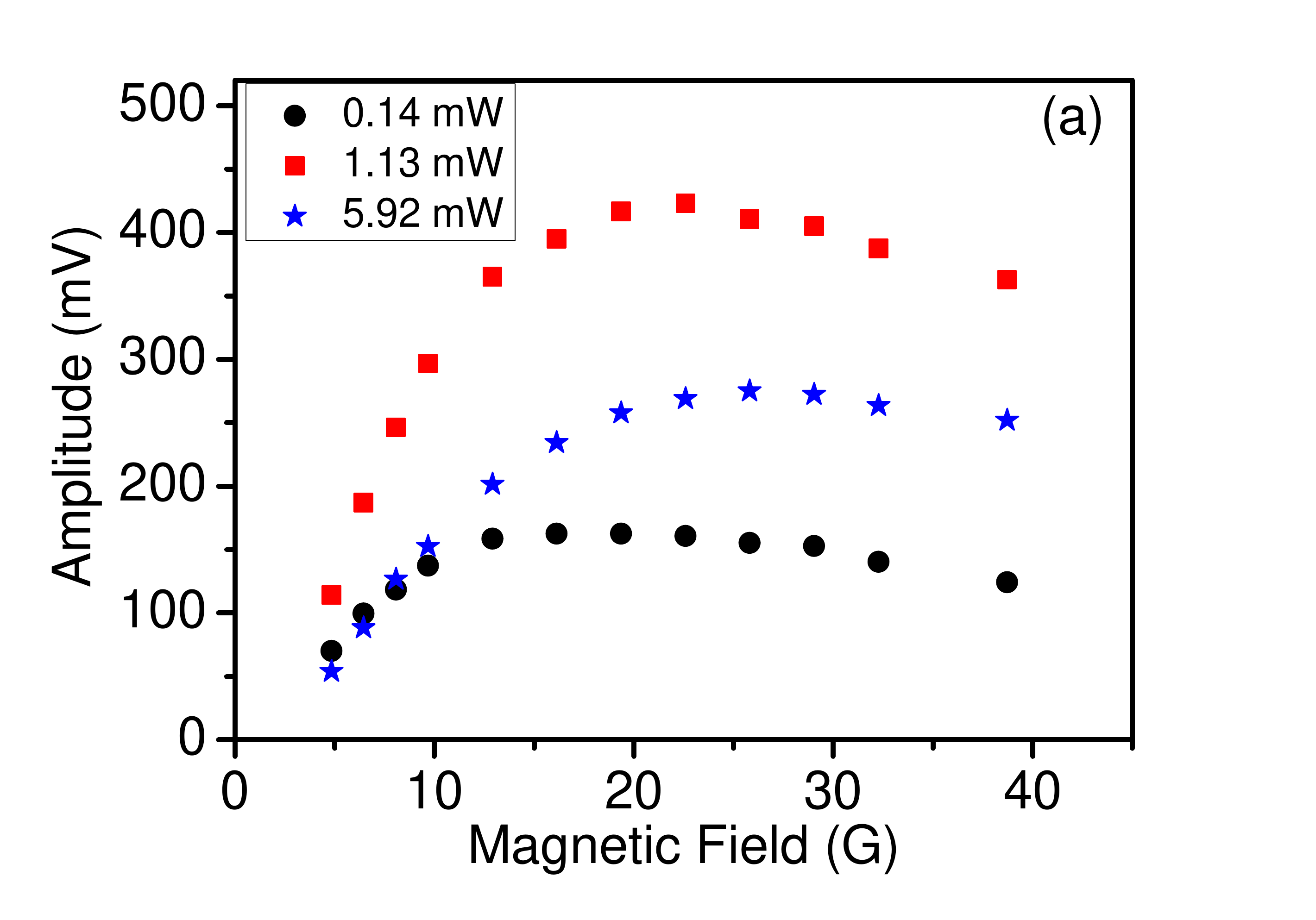}
\end{minipage}\qquad
\begin{minipage}[b]{.4\textwidth}
\includegraphics[width=1.2\linewidth]{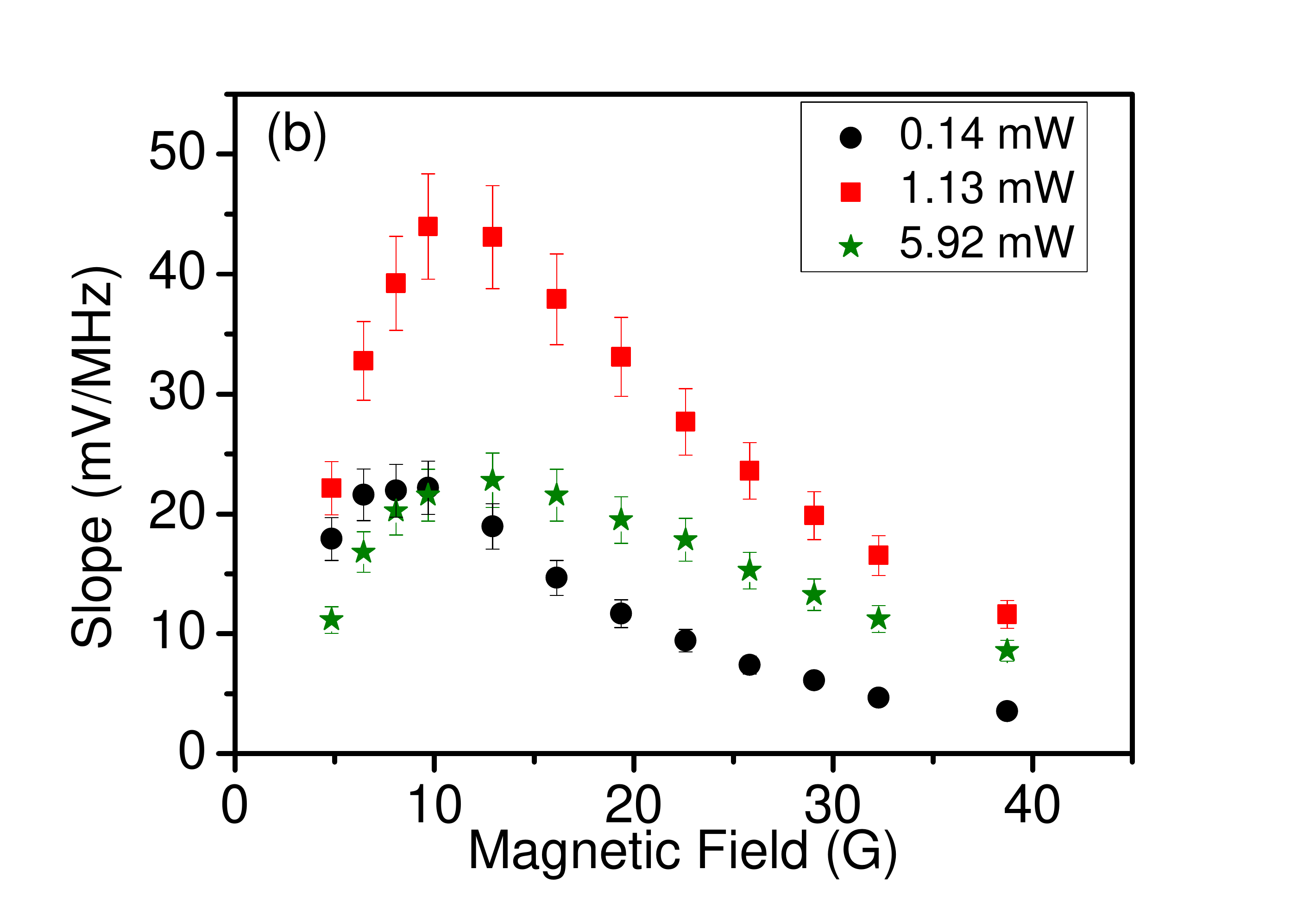}
\end{minipage}
\caption{Dependence of (a) amplitude and (b) slope of the TDFDL signal on magnetic field for $F=2\rightarrow F'=3$ transition of $^{87}Rb$ atom. The data is recorded for probe laser beam power of 140 $\mu$W and for different pump laser beam power. TDFDL signal is 12 MHz red detuned from $F=2\rightarrow F'=3$ transition of $^{87}Rb$. }
\end{figure}

Fig. 6 shows the effect of magnetic field on (a) the amplitude and (b) slope of TDFDL signal for different values of the pump beam power. The TDFDL signal is 12 MHz red detuned from $F=2\rightarrow F'=3$ transition of $^{87}Rb$. The amplitude of the signal increases up to a field of 20 G and then starts to decrease. The slope of the signal initially increases with magnetic field and then achieves its maximum value of $\sim$ 43 mV/MHz for the pump power of 1.13 mW and applied magnetic field of $\sim$ 9.7 Gauss. A reduction in slope is observed for  magnetic fields larger than $\sim$ 9.7 G as the separation between $\sigma_{+}$ and $\sigma_{-}$ transitions for the probe absorptions becomes more than a few natural widths. Therefore, it is evident from the graph that optimised slope and a large amplitude of TDFDL signal is obtained for the magnetic field of $\sim$ 9.7 G and the pump beam power of 1.13 mW. The slope and the amplitude reduces significantly due to power broadening at the pump beam power greater than 1.13 mW. The behaviour of calculated TDFDL signal's amplitude and slope with magnetic field is similar to the experimental results. However, the optimum value of magnetic field ($\sim$9.7 Gauss) for the experimental TDFDL signal for $F=2\rightarrow F'=3$ transition of $^{87}Rb$ is different from the value for the calculated signal as shown in Fig. 3(b). This may be due to the reason that our calculations are based on simple two level model whereas the real atomic system consists of multiple sub-levels in ground as well as excited state. \\ 

\subsection{Frequency stabilization of laser using TDFDL signal and its application to Rb-MOT setup}
\label{sec:3}
The frequency stability of an ECDL laser system using TDFDL locking signal has been studied. A standard PID (Toptica, Germany) controller is used to lock the laser frequency. The TDFDL signal was fed to this PID controller to lock the laser frequency. The desired frequency was selected by setting the lock reference level equal to the signal level at that frequency. The laser system remained frequency locked for several hours during the course of experiments. The error signal after  frequency locking is recorded on a calibrated photo-diode. \\
\begin{figure*}
\centering
\includegraphics[width=8.0 cm]{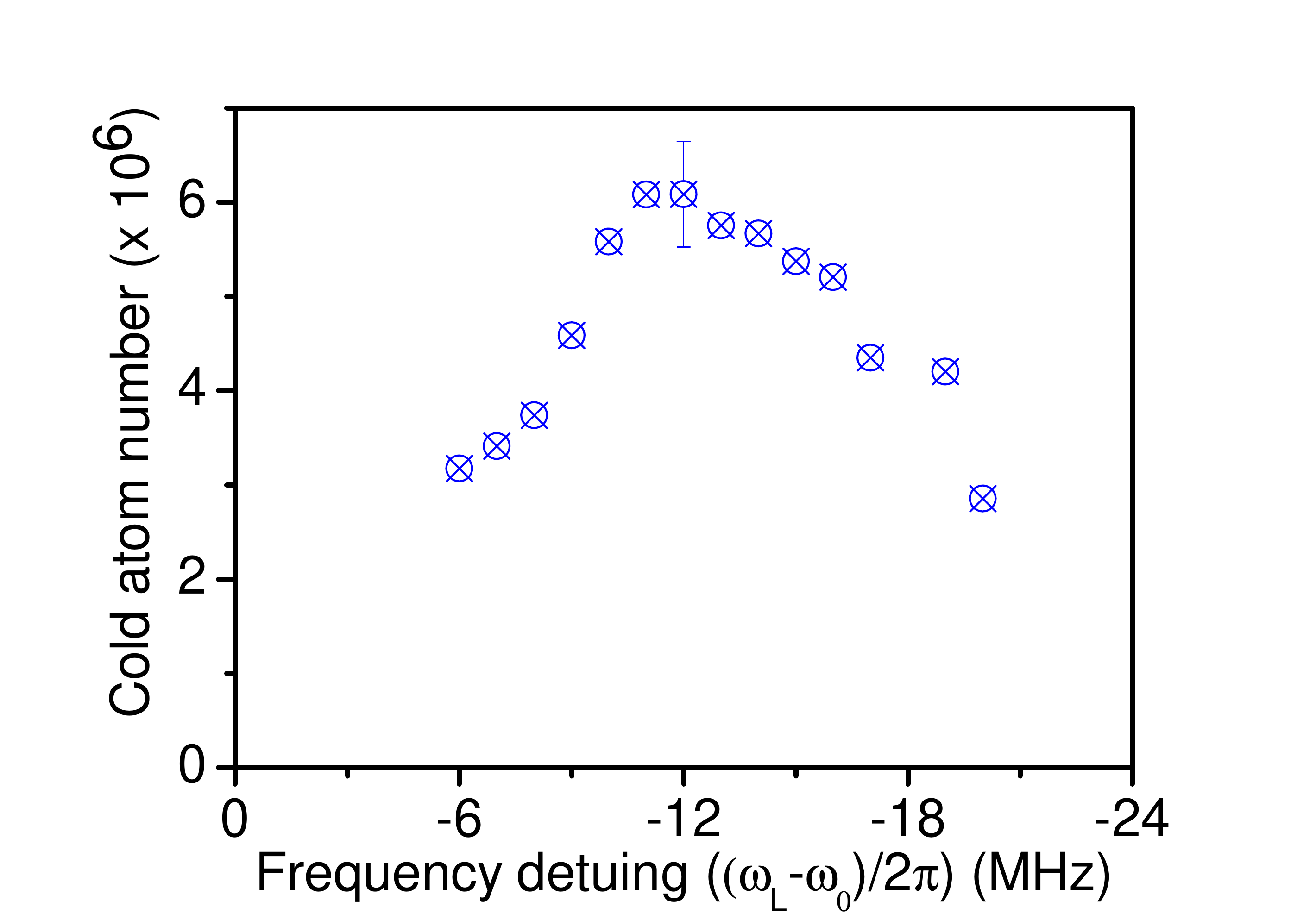}
\vspace*{0.1cm}       
\caption{The variation in maximum number of cold atoms in $^{87}Rb$-MOT with detuning of cooling laser frequency using TDFDL locking method. The MOT is pulsed loaded using Rb dispenser with dispenser current of 7 A for 5 second duration.}
\label{fig:7}       
\end{figure*}
Our proposed TDFDL technique has two advantages. First is robust locking due to use of a dispersive like signal with large slope, and second is large tunability without compromising the slope. The TDFDL signal has been used to lock the cooling laser frequency in the magneto-optical trap (MOT) setup. We can tune this cooling laser in locked condition itself, to precisely control the number of cold atoms in the MOT. The $^{87}Rb$ MOT is prepared in an octagonal chamber which is made of stainless steel with a residual pressure $\sim$ $1.4\times10^{-8}$ torr. The dispenser used in this experiment is a SAES Rb metal dispenser. Our dispenser is located at $\sim$ 19 cm form the MOT centre. Two feed-through pins on which dispenser is mounted are connected to external current supply from outside the chamber. Six laser beams of $\sim$ 5 mW power in each beam is used to form the MOT. The beam waist of each cooling laser beam is about $\sim$ 6 mm. The laser beam of an additional diode laser is combined with that of the cooling and trapping laser for re-pumping of the atoms. The quadrupole magnetic field gradient in axial direction was $\sim$ 10 G/cm. The MOT was operated in pulsed mode. In this, a current of $\sim$ 7 A is passed through Rb dispenser for 5 second duration and cooling laser frequency is changed using TDFDL setup. Fluorescence signal of pulsed loaded MOT is recorded on a calibrated photo-diode to measure the number in the MOT. In Fig. 7, the peak value of the number of atoms is plotted as a function of cooling laser detuning. The temperature of Rb dispenser is kept constant between each pulsed operation by flowing 1.5 A current to the dispenser. We have maintained $\sim$ 5 minute gap between each data. The precise control over laser frequency tuning using the TDFDL signal allowed the cooling laser detuning to be changed in steps of 1 MHz. We also note that the tuning was possible without breaking the frequency lock over a wide range. The cold atom number was found to be maximum ($\sim$ $6.0\times10^{6}$) at - 12 MHz of cooling laser detuning. The atom cloud temperature is typically $\sim$ 300 $\mu K$ for the values of cooling beam intensity and detuning used \cite{VB}. The result is in good agreement with the earlier work \cite{Lind, Arnold}.\\
\begin{figure*}
\centering
\includegraphics[width=8.0 cm]{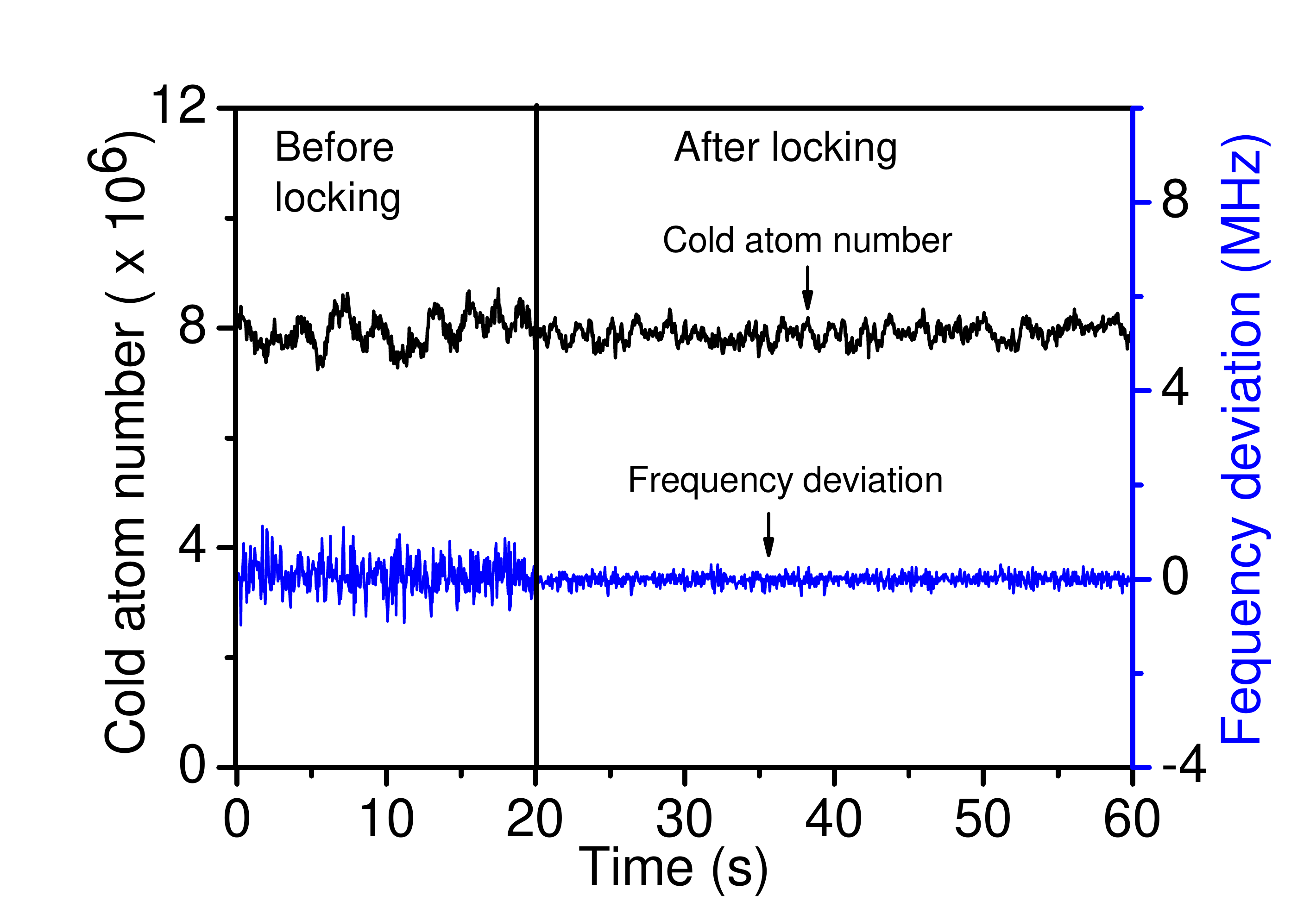}
\vspace*{0.1cm}       
\caption{The measured variation in number of cold atoms in the MOT and cooling laser frequency with time, before and after locking the laser frequency using TDFDL signal. The data is recorded with probe laser beam power of $\sim$ 140 $\mu$W and with pump laser beam power of $\sim$ 1.13 mW. The slope of TDFDL signal is $\sim$ 43 mV/MHz. The TDFDL signal was shifted $\sim$ - 12 MHz with respect to peak of the cooling transition of $^{87}Rb$. For these measurements, a continuous MOT was operated by applying a dc current of $\sim$ 2.7 A in the Rb dispenser.}
\label{fig:8}       
\end{figure*}
Fig. 8 shows the fluctuation in cooling laser frequency and the number of cold atoms in the MOT before and after locking the laser frequency. It is evident from this figure that both the number of cold atoms and error signal (i.e. frequency deviation) show the significant reduction in fluctuation after locking the laser frequency. Fig. 8 shows that fluctuation in number of cold atoms in the MOT gets reduced from $\sim$ $\pm$ 10 $\%$ (before locking) to $\sim$ $\pm$ 4.5 $\%$ (after locking) whereas the fluctuation in laser frequency gets reduced from $\sim$ $\pm$ 1 MHz (before locking) to $\sim$ $\pm$ 0.25 MHz (after locking). The laser is locked for more than 1 hour.\\  
\section{Conclusion}
\label{con}
In conclusion, a new laser frequency stablization technique which we call as tunable Doppler free dichroic lock (TDFDL) spectroscopy has been demonstrated. This technique provides a sharp and widely tunable dispersion like signal useful for frequency locking of a laser. The amplitude and the slope of the signal depends on the pump beam power and the applied magnetic field. Using the TDFDL signal, the tuning of zero-crossing point in the TDFDL locking signal is achieved by varying the relative frequency difference between the pump and probe beams. The TDFDL signal has been utilized for frequency locking and tuning of a laser used for cooling of $^{87}Rb$ atoms in a magneto optical trap.  
\section{Acknowledgements}
We are grateful to Amit Chaudhary for his help during the experimental work.

\end{document}